\documentclass[aps,prc,twocolumn,superscriptaddress,groupedaddress]{revtex4}  
\usepackage{graphicx}  
\usepackage{dcolumn}   
\usepackage{bm}        
\usepackage{amssymb}   
\usepackage{amsmath}

\begin{document}

\title{Insights into nuclear saturation density from parity violating electron scattering}
\author{C. J. Horowitz} \email{horowit@indiana.edu}      
\affiliation{Center for the Exploration of Energy and Matter and Department of Physics, Indiana University, Bloomington, IN 47405, USA}                             
\author{J. Piekarewicz}\email{jpiekarewicz@fsu.edu}
\affiliation{Department of Physics, Florida State University, Tallahassee, FL 32306, USA}              
\author{Brendan Reed} \email{reedbr@iu.edu}
\affiliation{Center for the Exploration of Energy and Matter and Department of Physics, Indiana University, Bloomington, IN 47405, USA}    
\affiliation{Department of Astronomy, Indiana University, Bloomington, Indiana 47405, USA}
\date{\today}

\begin{abstract}
\noindent
The saturation density of nuclear matter $\rho_0$ is a fundamental nuclear physics property that 
is difficult to predict from fundamental principles.  The saturation density is closely related to 
the interior density of a heavy nucleus, such as $^{208}$Pb.  Parity violating electron scattering 
can determine the average interior weak charge and baryon densities in $^{208}$Pb.  This requires 
not only measuring the weak radius $R_{\rm wk}$ but also determining the surface thickness of 
the weak charge density $a$.  We use the PREX experimental result for the weak radius of Pb and assume a 10\% theoretical uncertainty in the presently unmeasured surface thickness to obtain $\rho_0\!=\!0.150\pm0.010$ fm$^{-3}$.  Here the 7\% 
error also has contributions from the extrapolation to infinite nuclear matter.  These errors 
can be improved with the upcoming PREX II results and with a new parity violating electron 
scattering experiment, at a somewhat higher momentum transfer, to determine $a$. 
\end{abstract}

\maketitle

\section{Introduction}
The saturation density of nuclear matter $\rho_0$ is very important for the structure of nuclei.  Infinite nuclear matter, a hypothetical 
uniform system of protons and neutrons without Coulomb interactions, is expected to have an 
energy per nucleon that is minimized at $\rho_0$.  This minimum describes nuclear saturation 
and is a fundamental nuclear-structure property.  Furthermore, this value of $\rho_0$ is an important benchmark that is used to measure even higher density matter in astrophysics and in the laboratory.  Nuclear saturation implies that the interior density 
of heavy nuclei should be nearly constant and close to $\rho_0$.  Historically, the semi-empirical 
mass formula\,\cite{semiemp,semiemp1} and the liquid drop model\,\cite{liquiddrop} describe the
nucleus as an incompressible quantum drop at $\rho_0$.
But why does nuclear matter saturate? And how can one calculate the saturation density $\rho_0$?  Surprisingly, the answers to these deceptively simple questions have proved to be both subtle and elusive. 

Liquid water saturates at a density of 1\,g/cm$^3$ because of the size of the water molecules.  
Does nuclear matter saturate because of the finite nucleon size and if so, does this size explain the value of $\rho_0\!\approx\!0.15\,{\rm fm}^{-3}$\,?  The situation is likely more complicated. 
Nucleons are known to have repulsive cores because phase shifts for nucleon-nucleon scattering 
become negative at high energies (see http://nn-online.org). However, the core size is too small to 
explain the value of $\rho_0$\,\cite{bethe71}.  Indeed, nuclear matter calculations with 
only two-nucleon interactions may saturate at up to twice the expected density\,\cite{benday}.  It 
is now believed that three- and higher-nucleon interactions are important for nuclear saturation 
and for determining $\rho_0$.      

Chiral effective field theory (CEFT) provides a systematic expansion of the strong interaction 
between nucleons in powers of the momentum transfer over a suitable chiral 
scale\,\cite{CEFT1,CEFT0,CEFTreview}. This allows one to calculate the energy of nuclear 
matter to a given order in a chiral expansion. Note that CEFT includes two-, three-, and 
many-nucleon interactions. Under this framework, the empirical saturation point (density and 
energy per nucleon) are well reproduced within statistical and systematic 
uncertainties\,\cite{CEFT,CEFTlosalamos}.  The uncertainty band comes from the truncation 
of the chiral expansion and from imposing a cutoff at high momentum 
transfers. Whereas CEFT appears consistent with nuclear saturation at $\rho_0$, the error band in present calculations is too broad to make a sharp prediction of the actual value of $\rho_0$.

So if one can not accurately compute $\rho_0$ from first-principle calculations, can one observe it?  Strictly speaking, nuclear matter is an infinite system without Coulomb interactions, so observations of $\rho_0$ must involve an extrapolation from measurements in finite nuclei; see for example\,\cite{leptodermos}.  Nevertheless, the interior baryon density of heavy nuclei is expected to be fairly constant and close to $\rho_0$.  Among heavy nuclei, $^{208}$Pb may be particularly important because it is the heaviest stable doubly-magic nucleus.  As such, \emph{the interior baryon density of $^{208}$Pb may provide the finite nucleus observable that is most closely related to $\rho_0$.}  In this paper we present a new measurement of the interior baryon density of $^{208}$Pb based on results from the PREX experiment \cite{20,22}.  

Unfortunately, we do not have detailed knowledge of the neutron density in $^{208}$Pb; see Ref.\,\cite{tt} 
and references contained therein. The charge density is well measured so the proton density is accurately 
known\,\cite{1}.  However, $^{208}$Pb has 44 excess neutrons, so the neutron density can be significantly 
different from the proton density. Given this incomplete information, our present best estimate of $\rho_0$ 
comes from a variety of empirical nuclear energy density functionals. These functionals are calibrated to
the binding energies and charge radii of a variety of nuclei and can then be used to predict $\rho_0$, see 
for example Refs.\,\cite{mft,svmin}. In particular, Reinhard and Nazarewicz argue that fitting charge radii 
sharply constrains $\rho_0$\,\cite{witekrapid}.  

Alternatively, if one can cleanly measure the interior neutron density of $^{208}$Pb one should be able 
to infer $\rho_0$ with small and quantifiable uncertainties. Often neutron densities are determined with 
strongly interacting probes\,\cite{2}, such as antiprotons\,\cite{3,4}, elastic proton scattering\,\cite{5}, 
heavy-ion collisions\,\cite{7}, elastic pion scattering\,\cite{8}, and coherent pion photo production\,\cite{9}. 
One typically measures cross sections or spin observables that involve the convolution of the neutron density 
with an effective strong-interaction range for the probe. Although these observables can be measured 
with small statistical uncertainties, complexities arising from the strong interaction introduce significant 
systematic errors in the extracted neutron densities\,\cite{tt}.

It is also possible to measure neutron densities, or equivalently weak charge densities, with 
electroweak probes using coherent neutrino-nucleus scattering\,\cite{coherent,coherentAr,10,11} or 
parity violating (PV) electron scattering\,\cite{13,20}.  This is because the weak charge of a neutron is much 
larger than that of a proton, so the weak charge density of a nucleus is very closely related to its neutron 
distribution. Compared to strongly interacting probes, parity violation offers a clean and model-independent 
way to determine the weak charge density with much smaller uncertainties (statistical+systematic) than
with strongly interacting probes. In the last decades significant theoretical\,\cite{12,13,14,15,16,17,18,18a} 
and experimental\,\cite{19,20} efforts have been devoted to improve parity violating electron scattering 
experiments.  At Jefferson laboratory, the radius of the weak charge density of $^{208}$Pb was 
measured in the original PREX campaign\,\cite{20,22} and is now being measured with increased
precision during the follow-up PREX-II campaign\,\cite{PREXII}. At the same time, CREX will provide the first electroweak determination of the weak radius of $^{48}$Ca\,\cite{CREX}.

Present parity violating experiments focus on determining the rms radius of the weak charge 
density $R_{\rm wk}$ from a single measurement at a relatively low momentum transfer.  Yet additional 
features of the weak charge density $\rho_{\rm wk}(r)$ can be revealed by measuring the parity violating 
asymmetry $A_{pv}$ at higher momentum transfers. If $A_{pv}$ is measured at several momentum 
transfers, then a complete model independent representation of the weak charge density can be 
determined\,\cite{fullweak}, either as Fourier Bessel expansion or as a sum of Gaussians. This is 
feasible for $^{48}$Ca and may require measurements at six or seven momentum transfers.  
For $^{208}$Pb, however, this is more challenging because a determination of $\rho_{\rm wk}$ in 
the nuclear interior requires a measurement at high momentum transfer where the elastic cross 
section is very small.

What is then required to determine the saturation density $\rho_0$? In principle, one could follow
these four steps:  (a) Determine the entire weak charge density $\rho_{\rm wk}(r)$ of $^{208}$Pb; 
(b) average over $\rho_{\rm wk}(r)$ in the interior to obtain a measure of the average weak charge 
density; (c) combine this average weak charge density with an average of the experimental charge 
density to obtain a measure of the interior baryon density; (d) extrapolate such a value to the very 
closely related saturation density of infinite nuclear matter.  Here we combine the first two steps in 
a manner that dramatically minimizes the need for parity violating experiments.  

\section{Formalism}
We propose, rather than to determine the full density, a simple representation of $\rho_{\rm wk}(r)$ 
using a symmetrized two-parameter Fermi function that is then used to perform the interior average. 
That is, we model $\rho_{\rm wk}(r)$ as\,\cite{Sprung97,Piek16}
\begin{equation}
\rho_{\rm wk}(r,c,a)=\rho_{\rm wk}^{0} \frac{\sinh(c/a)}{\cosh{(r/a)}+\cosh(c/a)},
\label{eq.2pf} 
\end{equation}
where $c$ is the half-density radius, $a$ the surface diffuseness, and the normalization constant is
\begin{equation}
\rho_{\rm wk}^0=\frac{3Q_{\rm wk}}{4\pi c(c^2+\pi^2a^2)}
 \Rightarrow \int d^3 r \rho_{\rm wk}(r,c,a)=Q_{\rm wk}.
\label{eq.rho0}
\end{equation}
Here the total weak charge of a nucleus with $N$ neutrons and $Z$ protons is $Q_{\rm wk}=Q_nN+Q_pZ$,
where (including radiative corrections\,\cite{rad1,rad2}) $Q_n\!=\!-0.9878$ is the weak charge of a neutron,
and $Q_p\!=\!0.0721$ that of a proton. For $^{208}$Pb, $Q_{\rm wk}\!=\!-118.551$.

While the symmetrized Fermi (SFermi) function is practically indistinguishable from the conventional Fermi 
function, its superior analytic properties allows one to determine the form factor as well as all its moments 
in closed form\,\cite{Sprung97,Piek16}. In particular, the mean square weak radius is
\begin{equation}
R_{\rm wk}^2= \frac{1}{Q_{\rm wk}} \int r^2\!\rho_{\rm wk}(r) d^3r =  
  \frac{3}{5}c^{2} + \frac{7}{5}(\pi a)^{2}\,.
\label{eq.Rw2}
\end{equation}
We propose to use $\rho_{\rm wk}^0$ in Eq.(\ref{eq.rho0}) as the measure of the average interior weak charge 
density, which for clarity we rewrite in terms of the weak radius $R_{\rm wk}$ rather than $c$:
\begin{equation}
 \rho_{\rm wk}^0= \frac{27Q_{\rm wk}}{4\pi(5R_{\rm wk}^{2}-4\pi^{2}a^{2})\sqrt{15R_{\rm wk}^{2}-21\pi^{2}a^{2}}}.
 \label{eq.2pfR}
\end{equation}
Given that we are interested only in the average density $\rho_{\rm wk}^0$ rather than on the full density, 
PV experiments need only to determine the weak radius $R_{\rm wk}$ and the surface 
thickness $a$. The existing PREX and PREX II\,\cite{PREXII} measurements are primarily sensitive 
to $R_{\rm wk}$, so an additional PV experiment at a somewhat higher momentum transfer could 
determine $a$\,\cite{Piek16}. We will describe this experiment in a forthcoming paper.   


\begin{figure}[hbt]
 \centering
 \includegraphics[width=0.475\textwidth]{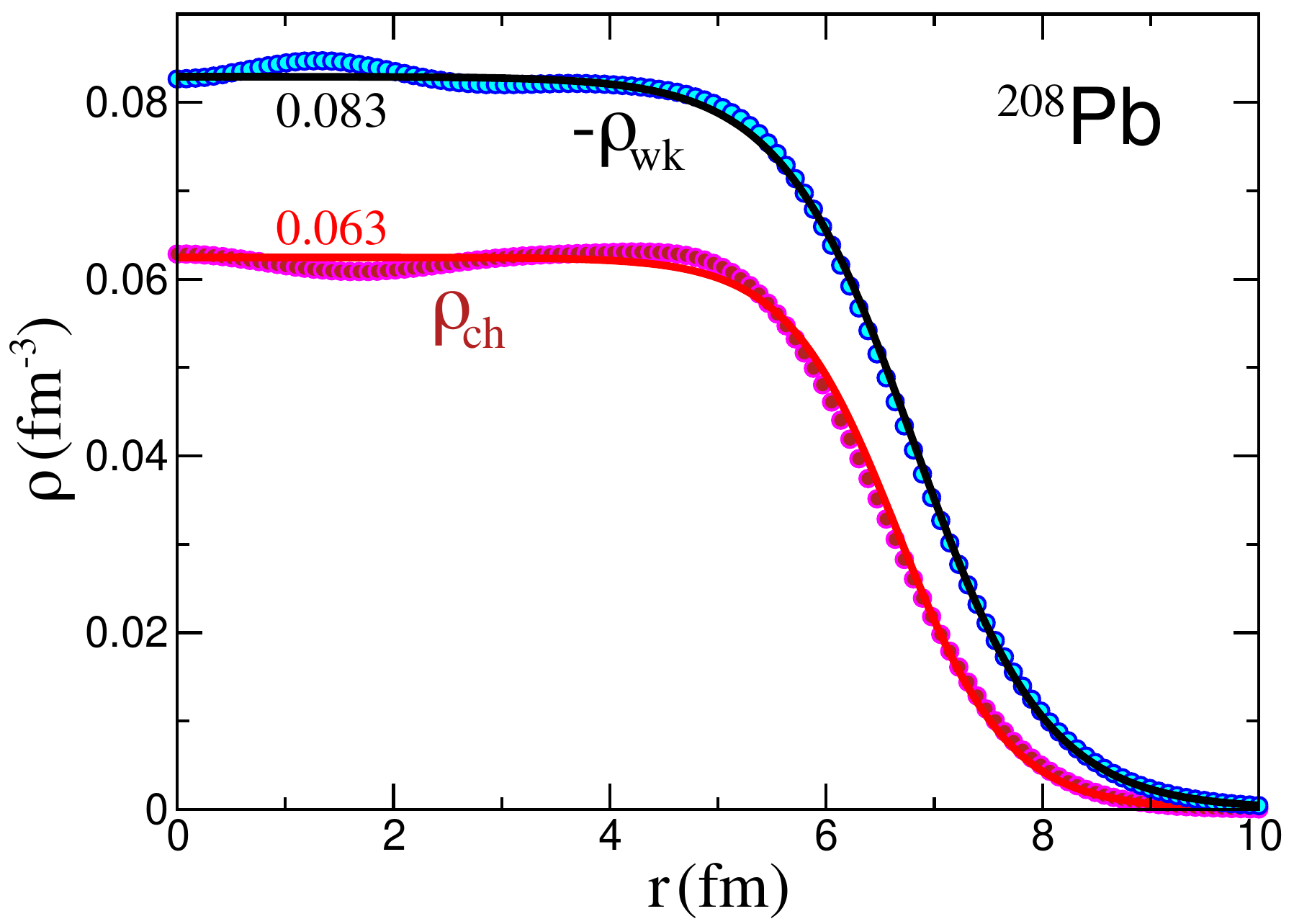}
 \caption{\label{fig:1} The experimental charge density of $^{208}$Pb\,\cite{1} (red circles) and the 
 corresponding SFermi function fit (solid red line).  Also shown is the weak charge density 
 as predicted by the FSUGold interaction\,\cite{FSUGold} (blue circles) along 
 with a SFermi function fit (solid black line).}
\end{figure} 

We illustrate our procedure in Fig.\,\ref{fig:1}, which shows the experimental charge density of $^{208}$Pb along 
with a SFermi function fit that yields: $c_{\rm ch}\!=\!6.6658$\,fm, $a_{\rm ch}\!=\!0.51219$\,fm, and a corresponding 
charge radius of $R_{\rm ch}\!=\!5.5031$\,fm\,\cite{1}. In turn, this implies a normalization of 
$\rho_{\rm ch}^0\!=\!0.06246$ fm$^{-3}$.  This is our measure of the average interior charge density of $^{208}$Pb.  
Figure \ref{fig:1} also shows a model weak charge density as predicted by the FSUGold relativistic mean field 
interaction\,\cite{FSUGold} and the corresponding SFermi function fit.  The SFermi functions---which average 
over shell oscillations---are seen to be very good representations of both the (electromagnetic) charge and weak 
charge densities.  Note that we are not proposing to use model predictions for the weak charge density but 
rather, a SFermi function with both parameters $R_{\rm wk}$ and $a$ determined from experiment. 
 
We now combine the average interior weak and charge densities to obtain an estimate of the average interior 
baryon density $\rho^0_{\rm b}$. That is,
\begin{align}
 \rho_{b}^0 & =  \rho_{n}^0 + \rho_{p}^0 
       = \frac{1}{Q_{n}}{\Big(\rho_{\rm wk}^0-Q_{p}\rho_{\rm ch}^0\Big)} + \rho_{\rm ch}^0 \nonumber \\
   & = \frac{1}{Q_{n}}\rho_{\rm wk}^0 + \left(1-\frac{Q_{p}}{Q_{n}}\right)\rho_{\rm ch}^0 \nonumber \\
   & = -(1.0123)\rho_{\rm wk}^0+(1.0730)\rho_{\rm ch}^0\, .
\label{eq.rhob0}
\end{align}
The final step is to extrapolate the interior baryon density $\rho_b^0$ to the closely related saturation density 
of infinite nuclear matter $\rho_0$.  {\it We define an extrapolation factor $f_{\rm ex}$ as the saturation density 
of infinite nuclear matter $\rho_0$ over the average interior density of $^{208}$Pb}:
\begin{equation}
f_{\rm ex}= \frac{\rho_0}{\rho_b^0}\, .
\label{eq.fex}
\end{equation}
We expect $f_{\rm ex}\!\approx\!1$.  
We estimate $f_{\rm ex}$ by considering a variety of relativistic and nonrelativistic energy density functionals (EDFs).  
For each EDF one calculates point proton $\rho_p(r)$ and neutron $\rho_n(r)$ densities and then computes the weak 
density by folding these point-nucleon densities with a dipole nucleon form factor of radius $r_{p}\!=\!0.84\,{\rm fm}$ 
that accounts for the finite nucleon size. Next, one fits SFermi functions to the model weak and charge densities to 
obtain $\rho_{\rm wk}^0$, $\rho_{\rm ch}^0$, and ultimately $\rho_b^0$ from Eq.(\ref{eq.rhob0}). Comparing this 
value of $\rho_b^0$ to the prediction for the saturation density $\rho_0$ yields $f_{\rm ex}$ for that particular EDF. 

\begin{figure}[hbt]
\includegraphics[width=.53\textwidth]{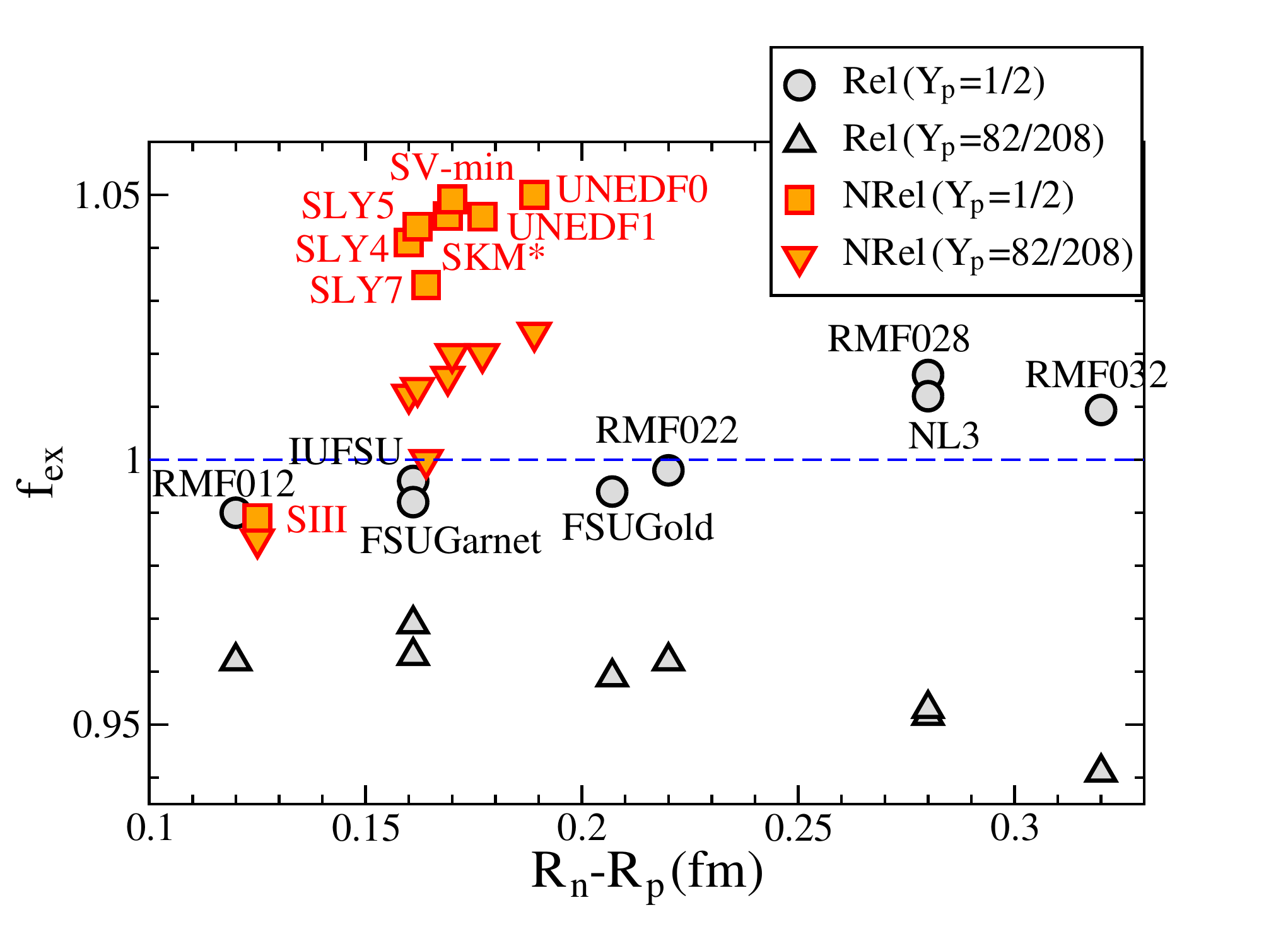}
\caption{\label{fig:2} The extrapolation factor $f_{\rm ex}$ defined in Eq.(\ref{eq.fex}) as a 
function of the the neutron skin thickness of $^{208}$Pb for a number of nonrelativistic 
and relativistic EDFs.  Shown with triangles is the extrapolation factor $\widetilde f_{\rm ex}$ 
to asymmetric nuclear matter with the same ratio of neutrons to protons as $^{208}$Pb. }
\end{figure}

Results are plotted in Fig.\,\ref{fig:2} for the following nonrelativistic Skyrme functionals: SIII\,\cite{SIII}, SLY4, SLY5, 
SLY7, and SKM*\,\cite{SLY7}, SV-min \cite{svmin}, UNEDF0 \cite{UNEDF}, and UNEDF1 \cite{UNEDF1}.  We also 
include results for the following relativistic functionals: FSUGold \cite{FSUGold},  IUFSU \cite{IUFSU}, NL3 \cite{NL3}, 
FSUGarnet, RMF012, 022, 028 and 032\,\cite{FSUGarnet}.
We see that $f_{\rm ex}$ is indeed close to one for all of the models that have been considered.  However, if one looks 
in more detail, $f_{\rm ex}$ for relativistic models is in general very close to one with a slight increase with increasing 
neutron skin (neutron minus proton radius $R_n\!-\!R_p$).  This is likely related to the density dependence of the 
symmetry energy which increases with increasing neutron skin. Most of the nonrelativistic models that we consider 
predict $f_{\rm ex}\!\approx\!1.04$ and this is noticeably larger than for the relativistic functionals. This is an interesting 
result that may be related to the assumed density dependence of the various EDFs.  For example, the old Skyrme 
force SIII, with $\gamma\!=\!1$ for the density dependent term $t_3\rho^{\gamma}$, predicts 
$f_{\rm ex}\!\approx\!0.99$ that is close to the prediction of most relativistic models. In contrast, all other Skyrme 
forces (shown in Fig. \ref{fig:2}) have smaller values for $\gamma$ and yield significantly larger $f_{\rm ex}$.  

The extrapolation from $\rho_b^0$ in $^{208}$Pb to $\rho_0$ involves three effects. First, surface tension---which is absent in an infinite system---increases the density of lead and tends to make $f_{\rm ex}\!<\!1$. Second, Coulomb interactions which are ignored in infinite nuclear matter reduce the density of lead making $f_{\rm ex}\!>\!1$.  
To some extent, effects from surface tension and Coulomb interaction cancel out restoring $f_{\rm ex}\!\approx\!1$. Finally, one is extrapolating in isospin from the neutron rich lead nucleus to symmetric nuclear matter, as $\rho_0$ is the saturation density of symmetric nuclear matter.

To explore the consequences of the extrapolation in isospin, we define $\widetilde\rho_0$ as the saturation density of asymmetric nuclear matter with a proton fraction identical to that of $^{208}$Pb, namely, 
$Y_p\!=\!82/208\!\simeq\!0.39$. It is a simple matter to calculate $\widetilde\rho_0$ for all EDFs included in Fig.\ref{fig:2}. Note that to a very good approximation $\widetilde\rho_0$ is given by\,\cite{Piek09}
\begin{equation}
 \frac{\widetilde\rho_0}{\rho_{0}} = 1 -3\frac{L}{K}\alpha^{2} + {\cal O}(\alpha^{4}),
 \quad \Big(\alpha\!=\!1\!-\!2Y_{p}\Big),
\label{eq.rho0tilde}
\end{equation}
where $K$ is the incompressibility coefficient of symmetric matter and $L$ the slope of the symmetry energy. 
Following Eq.(\ref{eq.fex}) we define in analogy $\widetilde f_{\rm ex}=\widetilde\rho_0/\rho_b^0$. Values for $\widetilde f_{\rm ex}$ are shown in Fig. \ref{fig:2} using up and down triangles. For relativistic functionals, $f_{\rm ex}\!\approx\!1$ and the interior density of lead is close to the saturation density of symmetric nuclear matter. However, $\widetilde f_{\rm ex}\!<\!1$ as $\widetilde\rho_0$ decreases with increases $L$, a quantity that is strongly correlated to $R_n\!-\!R_p$. In contrast, for nonrelativistic functionals $\widetilde f_{\rm ex}\!\approx\!1$ so the interior density of lead is close to the saturation density of asymmetric nuclear matter.    

This interesting difference between relativistic and nonrelativistic functionals should be explored using 
other models. For example, by building on $^{48}$Ca \cite{cc48Ca,NatureCa}, microscopic coupled 
cluster calculations for $^{208}$Pb may become feasible in the near future. This could provide a 
microscopic determination of $f_{\rm ex}$ that is more closely connected to chiral two- and three-nucleon 
forces.  Until then, we use all models in Fig.\,\ref{fig:2} to infer the following limit:  
\begin{equation}
f_{\rm ex}\approx 1.02 \pm 0.03\,.
\label{eq.fex1.02}
\end{equation}
That is, the extrapolation to infinite nuclear matter introduces a $\sim\!3\%$ uncertainty in the inferred 
value of $\rho_0$.


In summary, PV experiments can determine both the radius $R_{\rm wk}$ and surface 
thickness $a$ of the weak charge density of $^{208}$Pb, from which the average weak density 
$\rho_{\rm wk}^0$ is calculated using Eq.(\ref{eq.2pfR}). The known charge density 
$\rho_{\rm ch}^0$ is then added to $\rho_{\rm wk}^0$ in Eq.(\ref{eq.rhob0}) to obtain $\rho_b^0$.  
This, in turn, is extrapolated to $\rho_0$ using Eqs.(\ref{eq.fex}) and (\ref{eq.fex1.02}).

\section{Results}
We present a first estimate of $\rho_0$ based on the existing PREX result of 
$R_{\rm wk}\!=\!5.826\pm0.181$\,fm\,\cite{22}.  Unfortunately, at present there is no 
electroweak experiment that constrains the surface thickness $a$. Thus, we provide a 
conservative theoretical estimate for $a$. Considering all EDFs in Fig.\,\ref{fig:2} yields a 
surface thickness in the 0.58 fm (SIII) to 0.632 fm (RMF032) range. We arbitrarily select 
the UNEDF0 result to define the central value and assign a very conservative 10\% error 
that more than covers the theoretical range; that is, $a\!=\!0.616\pm0.062$\,fm. A
future PV experiment at a slightly larger momentum transfer to constrain $a$ would 
allow a direct experimental determination of the interior weak density.       

Adopting the PREX value for $R_{\rm wk}$, our theoretical assumption for $a$, 
and Eqs.(\ref{eq.2pfR}) and (\ref{eq.rhob0}) yields,
\begin{equation}
\rho_b^0=0.1473 \pm 0.0084 \pm 0.0030 \ {\rm fm^{-3}}\,,
\end{equation} 
were the first error is from the PREX error in $R_{\rm wk}$ while the second error 
corresponds to our assumed 10\% uncertainty in $a$.  The last step is to multiply 
this result by $f_{\rm ex}\!=\!1.02\pm0.03$ to get our present estimate for the 
saturation density of nuclear matter:
\begin{equation}
\rho_0=0.1502 \pm 0.0086 \pm 0.0031 \pm 0.0045 \ {\rm fm^{-3}},
\label{eq.rho0prex}
\end{equation}
where the last error is due to the uncertainty in $f_{\rm ex}$. Adding all three errors in 
quadrature gives a total uncertainty of 7\% that is dominated by the error in 
$R_{\rm wk}$. That is,  
\begin{equation}
\rho_0=0.150\pm 0.010\ {\rm fm}^{-3}\, .
\label{eq.rho0final}
\end{equation}
Our result is consistent, although somewhat lower, than the phenomenological 
estimate of $\rho_0\!=\!0.164\pm 0.007$\,fm$^{-3}$ claimed in Ref.\,\cite{CEFT} 
based on some selected density functionals---yet fully consistent with 
$\rho_0\!=\!0.151\pm 0.001$\,fm$^{-3}$ predicted by a relativistic EDF calibrated
using exclusively physical observables\,\cite{Piek014}. Note that an alternative 
procedure that uses a Helm-type\,\cite{22,helm} weak charge density instead of 
a SFermi function yields a consistent, yet slightly lower density than Eq.(\ref{eq.rho0final}).

\begin{figure}[htb]
 \centering
 \includegraphics[width=0.475\textwidth]{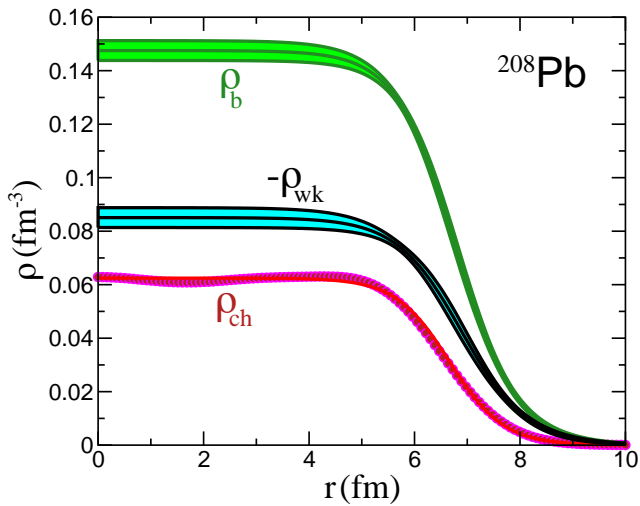}
 \caption{\label{fig:3} Theoretical prediction for the baryon density of $^{208}$Pb.  The error 
 band assumes that $R_{\rm wk}$ is measured to 1\% and the surface thickness is constrained to 
 10\%, see text for details. The corresponding curve for the weak charge density is also shown. 
 Finally, the experimental charge density\,\cite{1} is displayed along with a SFermi fit. }
\end{figure}

How accurately can $\rho_0$ be measured in the near future?  The PREX II 
campaign has completed data taking with a goal of measuring $R_{\rm wk}$ 
to 1\%.  Figure \ref{fig:3} shows an example baryon density for $^{208}$Pb 
assuming a SFermi weak charge density with $R_{\rm wk}\!=\!5.826$\,fm 
(central PREX value\,\cite{22}) and $a\!=\!0.62$ fm.  We have added the charge 
density as per Eq.(\ref{eq.rhob0}). The error band in Fig. \ref{fig:3} 
includes a 1\% error in $R_{\rm wk}$ and a 10\% error in 
$a$ added in quadrature.  This total error corresponds to $\pm 0.004$ fm$^{-3}$ 
in $\rho_b^0$ or about a 2.5\% error in $\rho_0$ that is comparable to our assumed 3\% error in $f_{\rm ex}$.      

There is strong motivation for an additional parity violating electron scattering experiment to measure the surface 
thickness $a$.  Both PREX and PREX II were performed at a momentum transfer of $q\approx 0.475$ fm$^{-1}$ 
and are primarily sensitive to the weak radius.  Instead, a new experiment near $q\approx 0.78$ fm$^{-1}$ 
is sensitive to $a$\,\cite{Piek16}.  Following\,\cite{big_paper} we have calculated the parity violating  
asymmetry $A_{pv}$ for elastic electron scattering including Coulomb distortions\,\cite{12}.  We find that the 
logarithmic derivative of $A_{pv}$ with respect to $\log(a)$ is about 0.53 at $q=0.78$ fm$^{-1}$. Therefore a 5\% 
measurement of $A_{pv}$ can constrain $a$ to 10\%.  We will discuss this possible experiment in more detail 
in a forthcoming paper.

\section{Conclusions}
In conclusion, the saturation density of nuclear matter $\rho_0$ is a fundamental nuclear physics property that 
is difficult to predict from chiral effective field theory.  Because of nuclear saturation, $\rho_0$ is closely related 
to the interior density of a heavy nucleus.   We emphasize that the average interior baryon density of $^{208}$Pb is an experimentally observable quantity that can be determined with parity violating electron scattering.  We used the existing PREX results for the weak radius to obtain a first measurement of the interior baryon density of $^{208}$Pb.  We then extrapolated 
this result to infinite nuclear matter and obtained $\rho_0\!=\!0.150\pm0.011$\,fm$^{-3}$.  The quoted 7\% error has contributions from the PREX error on the 
weak radius, uncertainty in a theoretical estimate of the surface thickness $a$, and the error in extrapolating to   
infinite nuclear matter. These errors can be improved with the upcoming PREX II results and with a new parity 
violating electron scattering experiment---at a somewhat higher momentum transfer---to determine the surface 
thickness of the weak density. This will allow an accurate determination of $\rho_0$ that is very closely related to the experimentally measured interior baryon density of $^{208}$Pb.   As a result of the parity violating measurements, the theoretical assumptions necessary to extract $\rho_0$ will be both reduced and clarified.

\section*{Acknowledgements}

We thank Witek Nazarewicz, Concettina Sfienti, Dick Furnstahl, and Zach Jaffe for helpful discussions.
This material is based upon work supported by the U.S. Department of Energy Office of Science, Office of Nuclear Physics under Awards DE-FG02-87ER40365 (Indiana University), DE-FG02-92ER40750 (Florida State University), and 
DE-SC0018083 (NUCLEI SciDAC Collaboration).


\end{document}